\documentclass[preprint,aip]{revtex4-1}

\usepackage{graphicx}
\usepackage{bm}%
\usepackage[colorlinks=true,linkcolor=blue]{hyperref}%
\expandafter\ifx\csname package@font\endcsname\relax\else
 \expandafter\expandafter
 \expandafter\usepackage
 \expandafter\expandafter
 \expandafter{\csname package@font\endcsname}%
\fi
\begin{document}

\title{Optical-resolution photoacoustic imaging through thick tissue with a thin capillary as a dual optical-in acoustic-out waveguide}%

\author{Olivier Simandoux}
\affiliation{ESPCI ParisTech, PSL Research University, CNRS, INSERM, Institut Langevin,1 rue Jussieu, 75005 Paris, France}%
\author{Nicolino Stasio}
\affiliation{Ecole Polytechnique F\'ed\'erale de Lausanne, Laboratory of Optics, CH-1015 Lausanne, Switzerland}
\author{J\'erome Gateau}
\affiliation{ESPCI ParisTech, PSL Research University, CNRS, INSERM, Institut Langevin,1 rue Jussieu, 75005 Paris, France}%
\author{Jean-Pierre Huignard}
\affiliation{Jphopto-consultant, 20 rue Campo Formio, 75013 Paris, France}%
\author{Christophe Moser}
\affiliation{Ecole Polytechnique F\'ed\'erale de Lausanne, Laboratory of Applied Photonics Devices, CH-1015 Lausanne, Switzerland}%
\author{Demetri Psaltis}
\affiliation{Ecole Polytechnique F\'ed\'erale de Lausanne, Laboratory of Optics, CH-1015 Lausanne, Switzerland}
\author{Emmanuel Bossy}
\email{emmanuel.bossy@espci.fr}
\affiliation{ESPCI ParisTech, PSL Research University, CNRS, INSERM, Institut Langevin,1 rue Jussieu, 75005 Paris, France}%

\begin{abstract}
We demonstrate the ability to guide high-frequency photoacoustic waves through thick tissue with a water-filled silica-capillary (150 $\mu$m inner diameter and 30 mm long). An optical-resolution photoacoustic image of a 30 $\mu$m diameter absorbing nylon thread was obtained by guiding the acoustic waves in the capillary through a 3 cm thick fat layer. The transmission loss through the capillary was about -20 dB, much lower than the -120 dB acoustic attenuation through the fat layer. The overwhelming acoustic attenuation of high-frequency acoustic waves by biological tissue can therefore be avoided by the use of a small footprint capillary acoustic waveguide for remote detection. We finally demonstrate that the capillary can be used as a dual optical-in acoustic-out waveguide, paving the way for the development of minimally invasive optical-resolution photoacoustic endoscopes free of any acoustic or optical elements at their imaging tip.
\end{abstract}


\maketitle

Biomedical photoacoustic imaging is a rapidly emerging imaging modality that has the unique capability to provide optical absorption contrast images at depth in tissue \citep{Beard2011,Wang2012,Ntziachristos2010}. It relies on the conversion of optical energy into acoustic waves. Amplitude modulated light, usually a laser pulse, is sent on the tissue of interest, penetrates the tissue and, as a consequence of the thermoelastic effect, photoacoustic waves are generated where light is absorbed. One of the advantages of photoacoustic imaging is that it can either provide label-free images of tissues based on the intrinsic variations of the absorption coefficient\citep{Zhang2006,Maslov2005}, or images based on exogenous contrast agents that can be used to target specific structures or metabolisms\citep{Agarwal2007,Li2008,Zerda2008}. 

In photoacoustic imaging there is an important trade-off between resolution and penetration depth\citep{Beard2011,Wang2012}. High resolution images require the detection of high frequency acoustic waves associated with high acoustic attenuation in tissue, thus limiting the distance between the absorbing structure and the ultrasound detection in high-resolution photoacoustic imaging. Photoacoustic imaging approaches can be divided in two categories based on the physical process that defines the image resolution\citep{Beard2011,Wang2012}. In acoustic-resolution photoacoustic imaging (AR-PAI) the resolution of the system is dictated by the highest detected frequency and aperture of the acoustic detection device. In optical-resolution photoacoustic imaging (OR-PAI), light is focused to a diffraction limited spot and scanned to illuminate point by point the region to be imaged\citep{Hu2009}. This yields optical resolution as the generation of photoacoustic waves is restricted to the illuminated region in the tissue. OR-PAI can be used as a valuable complement, or an alternative, to optical microcopies for shallow tissue imaging as it can specifically provide images of optical absorption in tissue with optical resolution. AR-PAI can provide images at much greater depths, up to a few centimeters, but only with acoustic resolution\citep{Beard2011,Wang2012}. OR-PAI relies on the ability to focus light into the tissue of interest and to detect the high frequency components generated by the small-sized illuminated region. Therefore it is limited by optical scattering to penetration depths smaller than 1 mm (typical value of the optical transport mean free path in tissue\citep{Ntziachristos2010}) and by the acoustic attenuation of high frequency waves.

At the cost of invasiveness, photoacoustic endoscopy allows imaging tissue at any depth by directly inserting an endoscopic device into tissue. Photoacoustic endoscopy was initially developed with acoustic resolution \citep{Viator2001,Sethuraman2007,Yang2009}, and various devices have been developed since then\citep{Jansen2014}.
Optical-resolution photoacoustic endoscopy (OR-PAE) was then first introduced by Zemp and coworkers, by use of a bundle of single-mode optical fibers\citep{Hajireza2011,Hajireza2013}, as in conventional optical endoscopy. The diameters of the experimental devices used typically ranged from 1 to a few millimeters. An alternative approach with a significantly smaller footprint than fiber bundles was recently proposed for optical endoscopy to address the issue of focusing and scanning light at depth in tissue, by use of thin  multimode fibers (typically 200 $\mu$m in diameter) and optical wavefront shaping, either via digital phase conjugation (DPC)\citep{Papadopoulos2012} or optical transmission matrix approaches\citep{Choi2012}. Optical phase conjugation focusing through a multimode fiber was subsequently applied for OR-PAE\citep{Papadopoulos2013}. Because OR-PAE generates photoacoustic waves with frequencies of typically several tens of MHz, the ability to detect externally the high frequency photoacoustic waves generated at depth in tissue is limited by the acoustic attenuation, typically 0.1-1.0 dB/cm/MHz\citep{Wells1999}.  In the two OR-PAE approaches introduced above\citep{Hajireza2013,Papadopoulos2013}, ultrasound was detected from outside either through a relatively thin tissue in the MHz range\citep{Hajireza2013}, or through a non-absorbing tissue phantom\citep{Papadopoulos2013}. Internal detection can be considered, but ultrasound sensors have to be miniaturized, resulting in limited sensitivity, and requiring dedicated technological developments\citep{Zhang2011,Miida2013}. No endoscopic device is currently able to achieve both optical scanning and focusing and detection of the generated photoacoustic waves to generate an optical-resolution photoacoustic image at centimeters depth in tissue.

\begin{figure}
\includegraphics[width=8.5 cm]{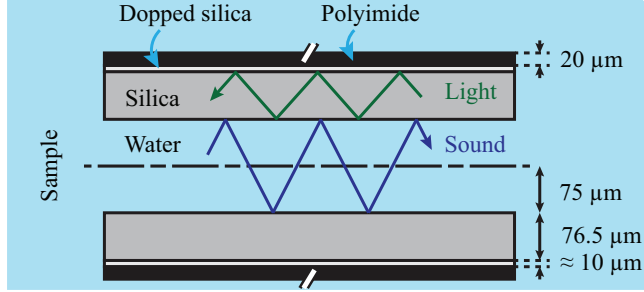}
\caption{Schematic of the water-filled capillary used as a dual optical-in acoustic-out waveguide. Light is guided in the silica cladding~\citep{SalehChap9} towards the image tip, while ultrasound is guided through the liquid core (the compressional and shear ultrasound velocity are respectively about $5970\ \mathrm{m.s^{-1}}$ and $3760\ \mathrm{m.s^{-1}}$, both much larger that the ultrasound velocity in water $\simeq 1490\ \mathrm{m.s^{-1}}$)~\citep{BlackstockChap12}. 
}
\end{figure}

In this work, we address the problem of detecting high frequency photoacoustic waves generated deep inside tissue for optical-resolution photoacoustic endoscopy. Classically, ultrasound detectors have been designed to detect the photoacoustic waves at the distal tip of minimally invasive devices inserted inside the tissue\citep{Zhang2011,Miida2013,Hajireza2013b}. Here we show that a water-filled capillary can be used to guide high frequency photoacoustic waves outside the tissue for remote ultrasound detection, thus avoiding absorption losses in tissue and the need for embedded miniaturized ultrasound detectors. We demonstrate in a second phase that the same capillary may also be used as an optical waveguide, and can therefore be used as a dual waveguide for both optical and acoustic waves. For both optics and acoustics, waves are guided into a core medium when the surrounding media have a lower index of refraction (or equivalently a higher phase velocity)~\citep{BlackstockChap12,SalehChap9}. A schematic of the capillary (Polymicro Technologies, USA) used for our study is shown in Fig. 1, illustrating waveguiding for both optics and acoustics. The silica tubing acts as a multimode waveguide for visible light, while the water-filled core acts as a waveguide for ultrasound.  The 150 $\mu$m inner diameter used in this work corresponds to an acoustic wavelength in water at 10 MHz. Therefore, for the ultrasound frequency range [5-30] MHz in this work, the acoustic waveguide can be considered a few-mode waveguide. In particular, the first-order quasi-piston mode exits the waveguide as a quasi-spherical wave, which can be detected by a spherically focused transducer.

\begin{figure}
\includegraphics[width=8.5 cm]{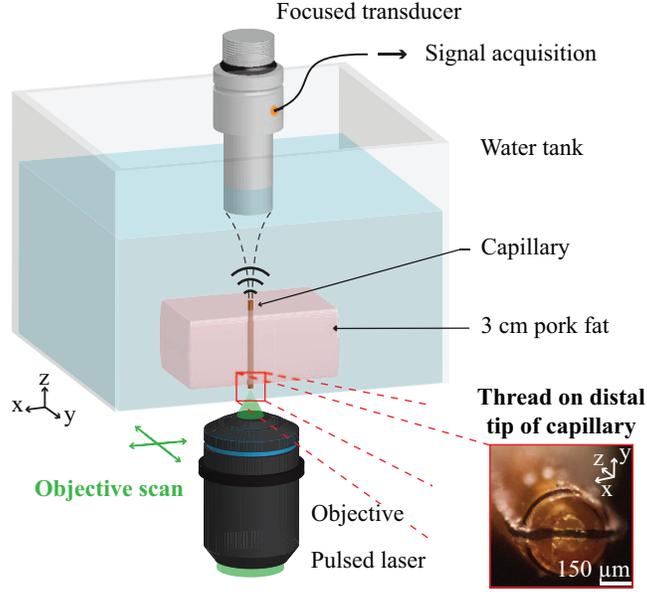}
\caption{Schematic of the experimental setup. The water-filled capillary is immersed in water and embedded in a pork fat layer. A microscope objective focuses and raster scans the optical spot around the distal tip of the capillary. The photoacoustic waves are transmitted through the hollow capillary and detected by a focused transducer. A 30 $\mu$m diameter black nylon thread, attached to the distal tip of the capillary, was used as absorbing sample (inset).}
\end{figure}

To demonstrate optical-resolution imaging with acoustic detection through a capillary, the experimental setup presented in Fig. 2 was used. In this case, the sample was directly illuminated at the imaging tip, without using the capillary as an optical waveguide. The output of a Q-switched pulsed laser (wavelength 532 nm, pulse duration 4 ns, repetition rate 10 Hz, Brilliant, Quantel, France) was focused by a 40X, 0.6 NA microscope objective (LUCPLFLN, Olympus, Japan), with a typical spot diameter around 5 $\mu$m.  The laser output energy was estimated around 1 $\mu$J/pulse. The sample (absorber + tissue layer), the ultrasound transducer and the capillary were immersed in a water tank. The objective was mounted on a two-axe motorized translation stage (M122.2DD, Physik Instrumente, Germany) installed below the water tank to mechanically raster scan the optical focus around the absorbing sample. A 30 mm long capillary was filled with water and was embedded in a 3 cm thick pork fat layer. The high acoustic attenuation of pork fat, and the significant thickness of the layer, were chosen to illustrate a situation in which the detection of high frequency photoacoustic signals at depth in tissue is highly challenging, if not impossible in practice. The distal tip, which corresponds to the entrance tip for the acoustic waves, exits the tissue close to the bottom of the water tank where the optical spot was raster scanned around the absorbing sample. The photoacoustic waves guided through the capillary fluid core (see Fig. 1) were detected at the proximal tip with a high frequency spherically focused transducer (20 MHz center frequency, 12.7 mm focal distance, 6.3 mm diameter, PI20-2-R0.50, Olympus, Japan). For each position of the focused optical spot, the photoacoustic signal was acquired by the transducer, amplified by a low noise amplifier (DPR500, remote pulser RP-L2, JSR Ultrasonics, USA), and digitized on an oscilloscope (DLM 2024, Yokogawa, Japan). The data was transferred to the computer used to control and synchronize the whole acquisition process.

We first estimated the transmission loss through pork fat by comparing photoacoustic signals propagating through water to those propagating through a 1-cm thick fat layer (limited by the 12.7 mm focal length of the transducer). For these measurements, a homogeneously absorbing 23 $\mu$m thick polyester red layer (color film 60193, R\'eflectiv, France) fixed at the bottom of the water tank and the transducer was placed so that the optical focus on the layer corresponds to the focal point of the transducer (no capillary was used). A typically 40-dB attenuation through the 1 cm fat layer was observed on the peak-to-peak amplitude.  The acoustic transmission through a 3 cm pork fat layer is therefore expected to be approximately 120 dB. The fat layer acted as a low-pass filter, as the acoustic attenuation increases with frequency. A spectral analysis showed that the central frequency of the signal is reduced from 17 MHz through water, close to the 20 MHz central frequency of the transducer, to only 5.5 MHz through the fat layer.
\begin{figure}
\includegraphics[width=8.5 cm]{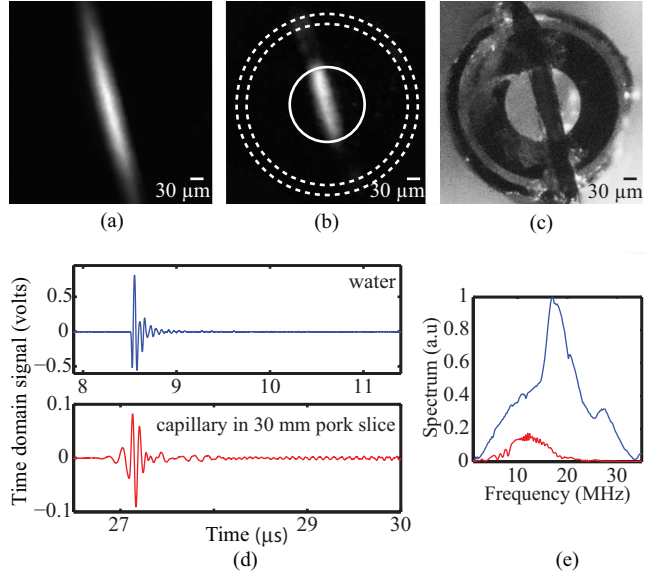}
\caption{\label{fig:images_and_signals}(a) Optical-resolution photoacoustic image of the thread obtained through water. (b) Optical-resolution photoacoustic image through a hollow capillary embedded in a 3 cm thick fat layer. The solid circle corresponds to the interfaces of the water-filled core, and the small and large dashed circles to the silica cladding and polyimide protective coating respectively. (c) White light optical image of the thread fixed at the distal tip of the hollow capillary. (d) Upper curve: typical time domain acoustic signal acquired through water only (average n= 128). Lower curve: typical time domain acoustic signal acquired through the capillary (average n= 1024). (e) Frequency analysis of the signals shown in (d).}
\end{figure}

We then demonstrated optical-resolution photoacoustic imaging with acoustic detection through a 3-cm thick pork layer by use of the capillary. A 30 $\mu$m diameter black nylon thread (NYL02DS, Vetsuture, France) was used as the sample to image (see Fig. 2). In Fig.~\ref{fig:images_and_signals}, the photoacoustic image and signal of the sample transmitted only in water, taken as references for our imaging system, were compared to the photoacoustic image and signal of the sample guided through the pork fat layer by the capillary. Optical-resolution photoacoustic images were obtained by raster scanning of the optical focus spot and measuring the peak amplitude of the envelope of the acoustic signal detected at each scanning position. The pixel size is 5 $\mu$m $\times$ 5 $\mu$m and fields of view are 400 $\mu$m $\times$ 400 $\mu$m. To obtain the reference image and signal, a thread was fixed at the bottom of the water tank in the focal region of the transducer, and no pork fat or hollow capillary was used. In figure~\ref{fig:images_and_signals}(a) we present the corresponding reference image of the thread. To form an image after propagation through the capillary across a 3 cm thick fat layer, the same absorbing thread was fixed at the distal tip of the hollow capillary (for easier alignment). The optical-resolution photoacoustic image obtained is presented in Fig.~\ref{fig:images_and_signals}(b) and closely resembles the white light optical image (Fig.~\ref{fig:images_and_signals}(c)): as expected, no loss of resolution is observed compared to the reference photoacoustic image in Fig.~\ref{fig:images_and_signals}(a) and the field of view is reduced to a 150 $\mu$m diameter circle, corresponding to the capillary core.
As illustrated by Fig.~\ref{fig:images_and_signals}(d) plotting the temporal signals and their corresponding frequency spectra, the acoustic coupling in and out of the capillary and acoustic propagation through the capillary act as a low-pass filter. Moreover, temporal spreading of the signal measured through the capillary indicates  modal dispersion. A typical -20 dB decrease of the peak-to-peak amplitude is observed through the capillary as compared with propagation in free water, which however remains several orders of magnitude smaller than the expected attenuation through the fat layer. This decrease can be attributed to both dispersion and absorption losses in the capillary. Preliminary experimental and theoretical results~\citep{SimandouxPhD2015} suggest that the main pulse in the capillary (observed at $t\simeq 27.2 \ \mu\mathrm{s}$) corresponds to a quasi-piston mode which attenuation is predominantly due to shear viscous loss inside the water-filled core, in addition to possible leaking of the guided wave through the silica walls. A systematic experimental and theoretical study of the acoustic propagation and transmission losses through the capillary is being carried out for a deeper understanding of the transmission efficiency of the capillary as an acoustic waveguide, but was beyond the scope of this letter.

\begin{figure}
\includegraphics[width=7.0 cm]{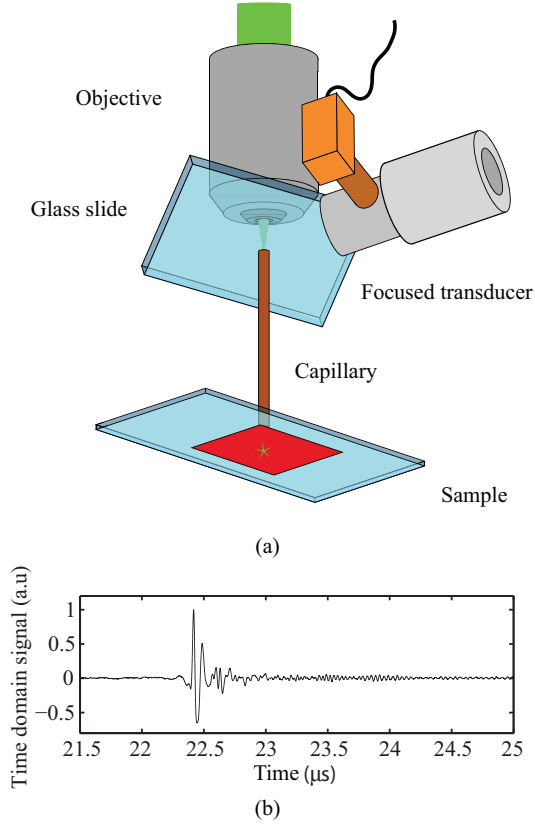}
\caption{\label{fig:dual_setup} (a) Proof-of-concept setup used to demonstrate photoacoustic sensing by use of the capillary as a dual waveguide. All the elements are immersed in water for acoustic coupling. Light is coupled into the capillary cladding through a $100\ \mu\mathrm{m}$-thick glass slide, which deflects the ultrasound guided through the water-filed cored towards the ultrasound transducer. The imaging tip (bottom tip) is free from any optical or acoustic elements. (b) Photoacoustic signal emitted from a homogeneously absorbing 23 $\mu$m thick polyester red layer placed at the imaging tip.}
\end{figure}

Finally, we demonstrated that the capillary may be used as a dual waveguide for both optical and acoustic waves, with an imaging tip free of any optical or acoustic elements, all deported to the tip outside the tissue. The corresponding experimental setup is shown in Fig~\ref{fig:dual_setup}.a. A transparent $100\ \mu\mathrm{m}$-thick glass slide was used to deflect the photoacoustic wave towards the ultrasound transducer oriented at $90^{\circ}$ from the capillary, while letting through the illumination pulse. A photoacoustic signal emitted by a homogeneously absorbing 23 $\mu$m thick polyester red layer is shown in Fig~\ref{fig:dual_setup}.b. It was verified that the light transmitted through the capillary was guided inside the silica cladding. The imaging resolution in this proof-of-concept configuration is therefore limited by the dimension of the silica tube. However, the silica cladding can also be used as a multi-mode optical waveguide through which light may be focused by wavefront shaping techniques, such as optical phase conjugation already demonstrated for optical-resolution by~\citet{Papadopoulos2013} in a multi-mode fiber.

In conclusion, our results therefore show that a fluid-fill silica capillary could be used as a dual mode waveguide (optical in, acoustic out) to perform optical-resolution photoacoustic imaging with an imaging tip free of any optical or acoustic elements (all being deported to the tip outside the tissue), and pave the way for minimally invasive optical-resolution photoacoustic endoscopy.\\

Olivier Simandoux gratefully acknowledges funding from the Direction G\'en\'erale de l'Armement (DGA). J\'erome Gateau gratefully acknowledges funding from the Plan Cancer 2009-2013 (Action 1.1, Gold Fever). The work at Institut Langevin was partially funded by the LABEX WIFI (Laboratory of Excellence ANR-10-LABX-24) within the French Program “Investments for the Future” under reference ANR-10- IDEX-0001-02 PSL*.

\bibliography{ORPAI_Capillary}

\end{document}